\documentclass[aps,prl,twocolumn,showpacs,groupedaddress]{revtex4}
\usepackage{amsmath,amssymb}
\usepackage{graphics,color,epsfig}

\begin{document}
\title{Inversionless light amplification and optical switching controlled by
state-dependent alignment of molecules}
\author{A.~K.~Popov} \altaffiliation [Corresponding author:]
{\,Alexander Popov}\email{apopov@uwsp.edu}
\homepage{http://www.kirensky.ru/popov} \affiliation{ Department
of Physics \& Astronomy and Department of Chemistry, University of
Wisconsin-Stevens Point, Stevens Point, WI 54481, USA}
\affiliation{Institute of Physics of the Russian Academy of
Sciences, 660036 Krasnoyarsk, Russia}
\author{V.~V.~Slabko}
\email{slabko@iph.krasn.ru} \affiliation{Krasnoyarsk State
Technical University, 660074 Krasnoyarsk,
Russia}\affiliation{Institute of Physics of the Russian Academy of
Sciences, 660036 Krasnoyarsk, Russia}
\date{October 6, 2003}
\begin{abstract}
We propose a method to achieve amplification without population
inversion by anisotropic molecules whose orientation by an
external electric field is state-dependent. It is based on
decoupling of the lower-state molecules from the resonant light
while the excited ones remain emitting. The suitable class of
molecules is discussed, the equation for the gain factor is
derived, and the magnitude of the inversionless amplification is
estimated for the typical experimental conditions. Such switching
of the sample from absorbing to amplifying via transparent state
is shown to be possible both with the aid of dc and ac control
electric fields.
\end{abstract}
 \maketitle
Amplification of light is determined by the difference between the
net absorbtion and stimulated emission of radiation. Usually, this
requires larger population of the upper quantum state then the
lower one that are coupled by the induced transitions. However,
the amount of absorbed and emitted photons depend not only on the
populations of the resonant energy levels but also on the
probabilities of the induced transitions and on distribution of
the populations over the energy-degenerated states. Therefore, the
inversion of population is, generally speaking, only a special
case, in which the amount of the emitted light prevails over the
absorbed one. Based on this fact, a variety of realizations of
amplification without population inversion (AWI) was proposed in
the early years of quantum electronics. Such possibility based on
a possible difference in velocity distribution in upper and lower
states was discussed in \cite{65}. Nonreciprocity of probabilities
of induced emission and absorbtion caused by nonlinear
interference affect in the field of the auxiliary radiation was
investigated for two-level atoms in \cite{RSob} and for
three-level systems in \cite{FeokSok}. Corresponding AWI at
transitions of neon was predicted and analyzed in details in
\cite{LWI} and proved in the experiments \cite{Bet}. The asymmetry
in the lineshape of net emission and absorption at the transitions
to the autoionizing states and related possibility of AWI was
considered in \cite{AG}. A possible AWI in dichroic molecules was
pointed out in \cite{PS}. The feasibility of AWI of short pulses
in three-level system was considered in \cite{KKh}. A review of
later publications on AWI is given in \cite{Corb}. An asymmetry in
absorption and emission lineshape of two-level system caused by
its interaction with a thermostat is discussed in recent
publication \cite{Sh}. In this paper we propose and discuss a
method of achieving lasing without population inversion by the
anisotropic molecules. It bases on selective alignment and
consequent decoupling the lower-state molecules from the polarized
resonant radiation, while the upper-state molecules remain
amplifying this radiation through its stimulated emission. 

In electrodipole approximation, the probability of induced
transitions between levels $m$ and $g$, $W_{mg}$, depends on the
factor $d_{mg}\cos \theta $, which is the projection of the
electrodipole transition matrix element on a direction of a vector
of a resonant oscillating electric field that causes such
transition,
\begin{equation}
W_{mg}={B}|E|^2F(\omega).\label{w}
\end{equation}
Here, ${B}=8\pi |d_{mg}\cos \theta|^2/\hbar^2$ is the Einstein
coefficient, $F(\omega)$ is the transition form-factor, $\omega$
is frequency of radiation. For molecules, the direction of
$\mathbf{d}_{mg}$ is stipulated by orientation of a molecule in
space and by its symmetry. Under influence of an external dc
electric field, a molecule turns towards the direction that
corresponds to the minimum of potential energy of their
interaction. Thus orientation of a molecule is determined by both
the direction of an external field and by the symmetry of a
molecule.  The degree of alignment of a molecular medium depends
on the alignment parameter, which is given by the ratio of the
interaction energy with the external field $U$ to the energy of a
thermal motion $ъв$ that renders disorientation. However, the
energy of such interaction $U$, and consequently the degree of
orientation, can be different for the molecules in lower and upper
states. Therefore, in such case the probabilities of induced
transitions corresponding to emission and absorption  of polarized
light, averaged over the molecules with different orientations,
are not equal. This opens an opportunity of AWI for the polarized
light through manipulating the difference in the orientation
degree of the molecules at the different energy levels by the
control field.

Consider amplification index $\alpha > 0$ (absorption, if $\alpha
< 0$), which determines exponential change of intensity for the
plain-polarized probe radiation $\mathbf{Х}$ along the medium $I =
I_0\exp\{\alpha z\}$. In the presence of the control field
$\mathbf{E}_0$, it is given by the equation
\begin{equation}
\alpha=\sigma_0\int [n_mf_m(\Theta, \mathbf{E}_0)-n_gf_g(\Theta,
\mathbf{ E}_0)]\cos^2 \theta d\Theta.\label{a}
\end{equation}
Here, $n_m$ and $n_g$ are populations of upper and lower levels
created by any common way, $d\Theta=\sin \theta d\theta d\phi$ is
element of a solid angle, and $\sigma_0=8|\pi d_{mg}|^2\omega
F(\omega)/c\hbar$ is absorption/emission crosssection for the
molecules, whose transition electrodipole moment is aligned along
the polarization of the probe field $\mathbf{E}$.

The functions $f_m(\Theta,\mathbf{E}_0)$ and
$f_g(\Theta,\mathbf{E}_0)$ describe the distribution of the
molecules over orientations at upper $m$ and lower $g$ energy
levels. They depend on the $U_{m,g}(\Theta,\mathbf{E}_0)$, the
energy of their interactions with the control field $\mathbf{E}_0$
in the corresponding states. If the energy level lifetime excesses
significantly  the time required to set the orientation balance,
they are given by the Boltzmann distribution
\begin{equation}
f_j(\Theta, \mathbf{E}_0)=A_j\exp\{-U_j
(\Theta,\mathbf{E}_0)/kT\}.\label{f}
\end{equation}
Here, $A_j^{-1}=\int\exp\{-U_j(\Theta,\mathbf{E}_0)/kT\}d\Theta$
is a scale factor, T - temperature, k -  Boltzmann constant, and
$j = \{g, m\}$. Potential energy of molecules coupled with the
electric field $\mathbf{E}_0$ can be written as follows \cite{Kie}
\begin{equation}
U_j (\Theta,\mathbf{E}_0)=
\mu_i^jE_{0i}-\beta_{ik}^jE_{0i}E_{0ik}.\label{u}
\end{equation}
Here, $\mu_i^j$ is i-component of a vector of the permanent dipole
moment and $\beta_{ik}^j$ is the component of the tensor of
electrical polarizability, both for the molecules at energy level
$j$. The first term in Eq. (\ref{u}) represents the energy of
molecules without center of symmetry, which possess a permanent
dipole moment. The second term depicts the interaction energy with
the dipole induced by the field $\mathbf{E}_0$.  This term
describes the alignment, which can be  also caused  by the strong
optical field with the amplitude $\mathbf{E}_0$ and frequency
$\omega_0$ \cite{Stap}.

Let's consider the example of axial-symmetric molecules and assume
the directions of both permanent $\mathbf\mu$ and of the
electrodipole transition moment aligned along the symmetry axis of
the molecule, which makes an angle $\theta_0$ with the control
field $\mathbf{E}_0$. Then the interaction energy (\ref{u}) and
distribution functions (\ref{f}) become dependent only on the
angle $\theta_0$
\begin{equation}
f_j(\theta_0, \mathbf{E}_0)=A_j\exp\{p_j\cos\theta_0\pm
q_j\cos^2\theta_0\}.\label{ft}
\end{equation}
Following \cite{Kie}, we introduce parameters of permanent dipole
orientation $p_j$ and of the alignment determined by the
polarizability ellipsoid $q_j$ as
\begin{equation}
p_j=\mu^jE_0/kT,\quad q_j=(b^j_{33}-b^j_{11})E^2_{0}/2kT.
\label{pq}
\end{equation}
Here, $b^j_{33}$ and $b^j_{11}$ are principal values of the
polarizability tensor along the symmetry axis and across it,
accordingly, for the molecule at level $j$. For the control ac
field $\mathbf{E}_0$ at $\omega_0$ the values $b_{33}$ and
$b_{11}$ are given by equation \cite{Kie}
\begin{equation}
b_j=(e^2/m)\sum_l d_{lj}/(\omega^2_{lj}-\omega^2_0), \label{b}
\end{equation}
where $m$ and $e$ are mass and charge of electron, $\omega_{lj}$
is frequency and $d_{lj}$ is corresponding projection of the
electrodipole moment of transition between levels $j$ and $l$.

\begin{figure}[!h]
\begin{center}
\includegraphics[width=.45\textwidth]{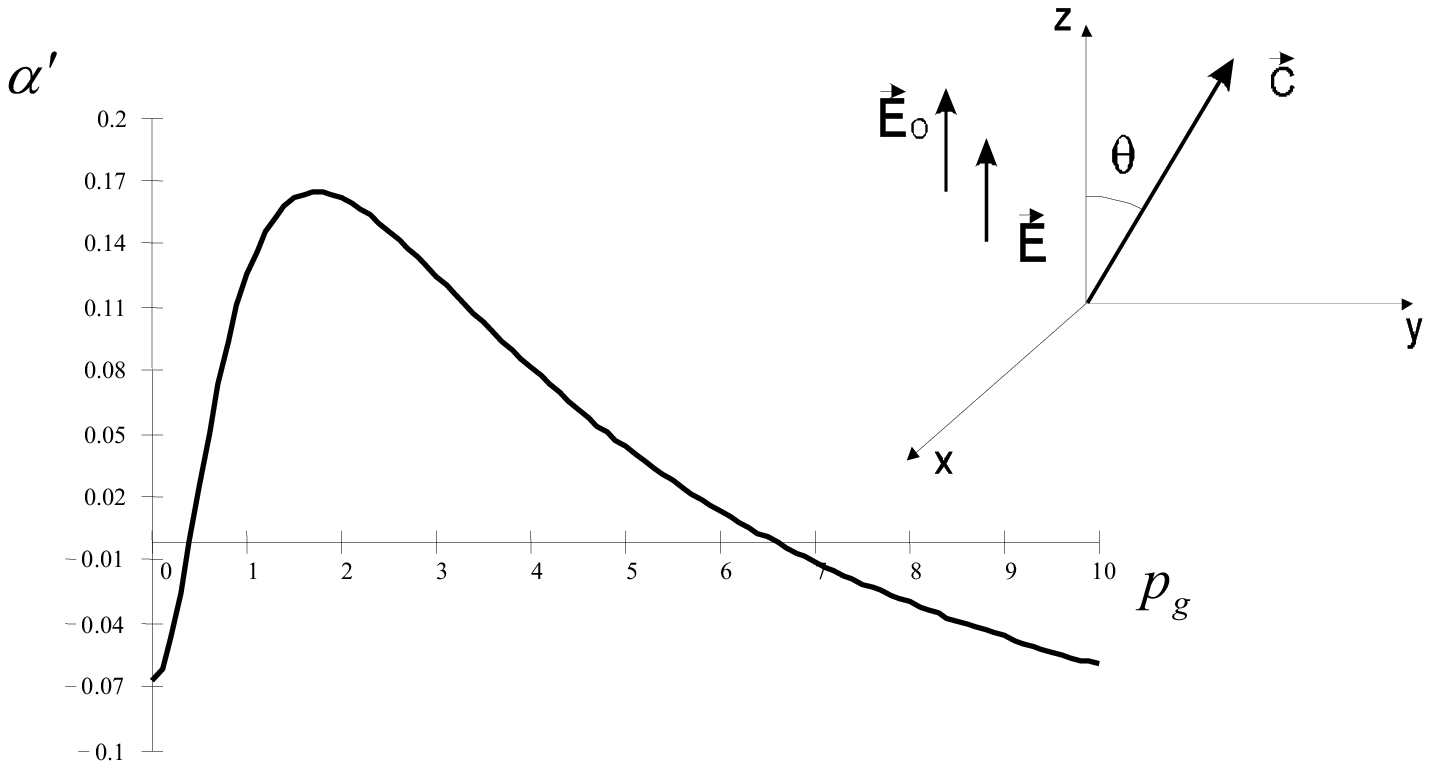}
\caption{\label{f1}Scaled amplification index
$\alpha'=\alpha/n_g\sigma_0$ vs alignment parameter
$p_g=\mu_gE_0/kT$ at $\mu_m/\mu_g$=4, $n_m/n_g$=0.8 ($\mathbf{E}
\upuparrows\mathbf{E}_0$). Orientations of the probe field
$\mathbf E$, molecule symmetry axis $\mathbf C$ and of the control
field $\mathbf E_0$ are depicted in the inset.}
\end{center}
\end{figure}

\begin{figure}[!h]
\begin{center}
\includegraphics[width=.45\textwidth]{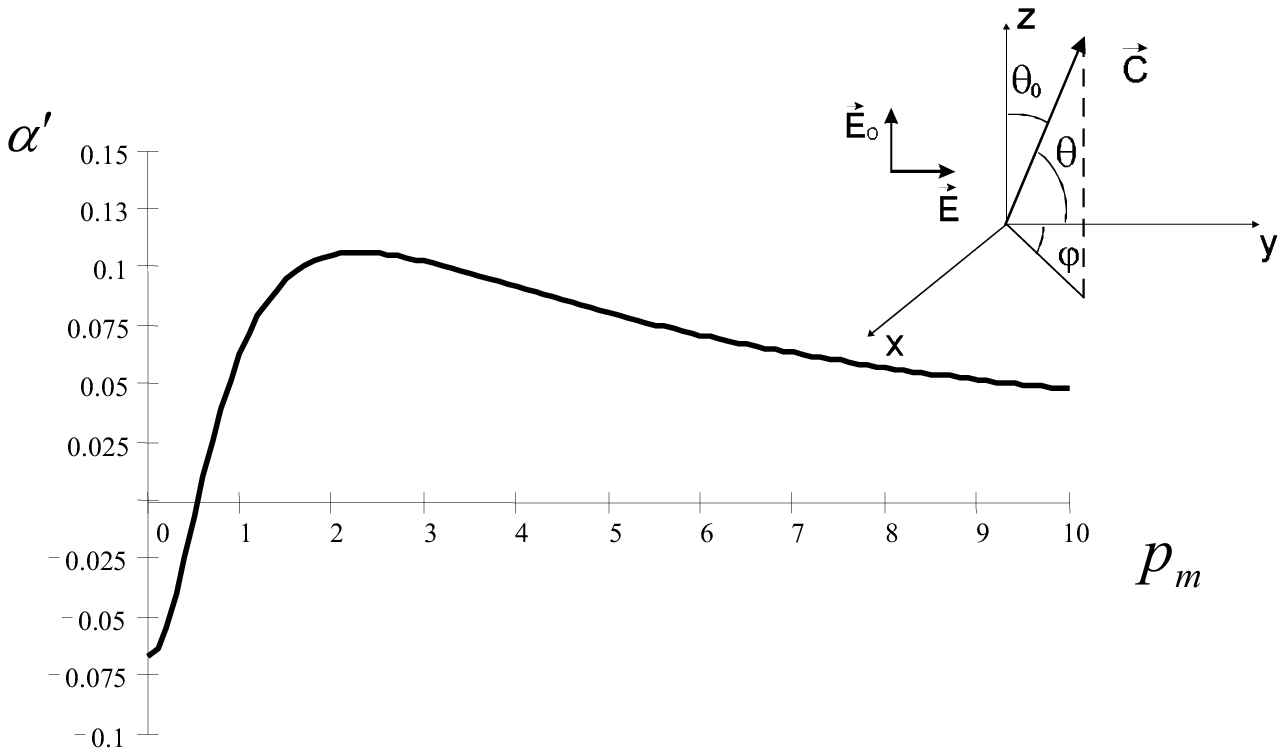}
\caption{\label{f3}Scaled amplification index
$\alpha'=\alpha/n_g\sigma_0$ vs alignment parameter
$p_m=\mu_mE_0/kT$ at $\mu_g/\mu_m$=4, $n_m/n_g$=0.8 ($\mathbf{E}
\perp\mathbf{E}_0$). Orientations of the probe field $\mathbf E$,
molecule symmetry axis $\mathbf C$ and of the control field
$\mathbf E_0$ are depicted in the inset.}
\end{center}
\end{figure}

We shall illustrate the effect of inversionless amplification for
two cases $\mathbf{E} \upuparrows\mathbf{E}_0$ and $\mathbf{E}
\perp\mathbf{E}_0$ depicted in Figs. \ref{f1} and \ref{f3}, where
the molecule symmetry axis $\mathbf{C}$, $\mathbf{\mu}_{m,g}$, and
$\mathbf{d_{mg}}$ coincide. If the permanent dipole moment in the
upper energy level is larger, than in the lower one
($\mu_m>\mu_g$), then  AWI occurs for $\mathbf{E}$ polarized along
$\mathbf{E}_0$. In this case $\theta=\theta_0$, and Eq. (\ref{a})
takes the form
\begin{equation}
\alpha=\sigma_0\left[n_m(1-\frac{2}{p_m})L(p_m)-n_g(1-\frac{2}{
p_g})L(p_g)\right],\label{a1}
\end{equation}
where  $L(p_j)$ is Langevin function \cite{Kie,Jahn}
\begin{align}
&L(p_j)=\coth
p_j-\frac{1}{p_j}=(1-\frac{2}{p_j})^{-1}\nonumber\\&\times\int^{2\pi}_0
d\varphi\int^{\pi}_0d\theta_0 A_j\exp\{p_j
\cos\theta_0\}\cos^2\theta_0\sin\theta_0.\label{ft}
\end{align}
\begin{figure}[!h]
\begin{center}
\includegraphics[width=.45\textwidth]{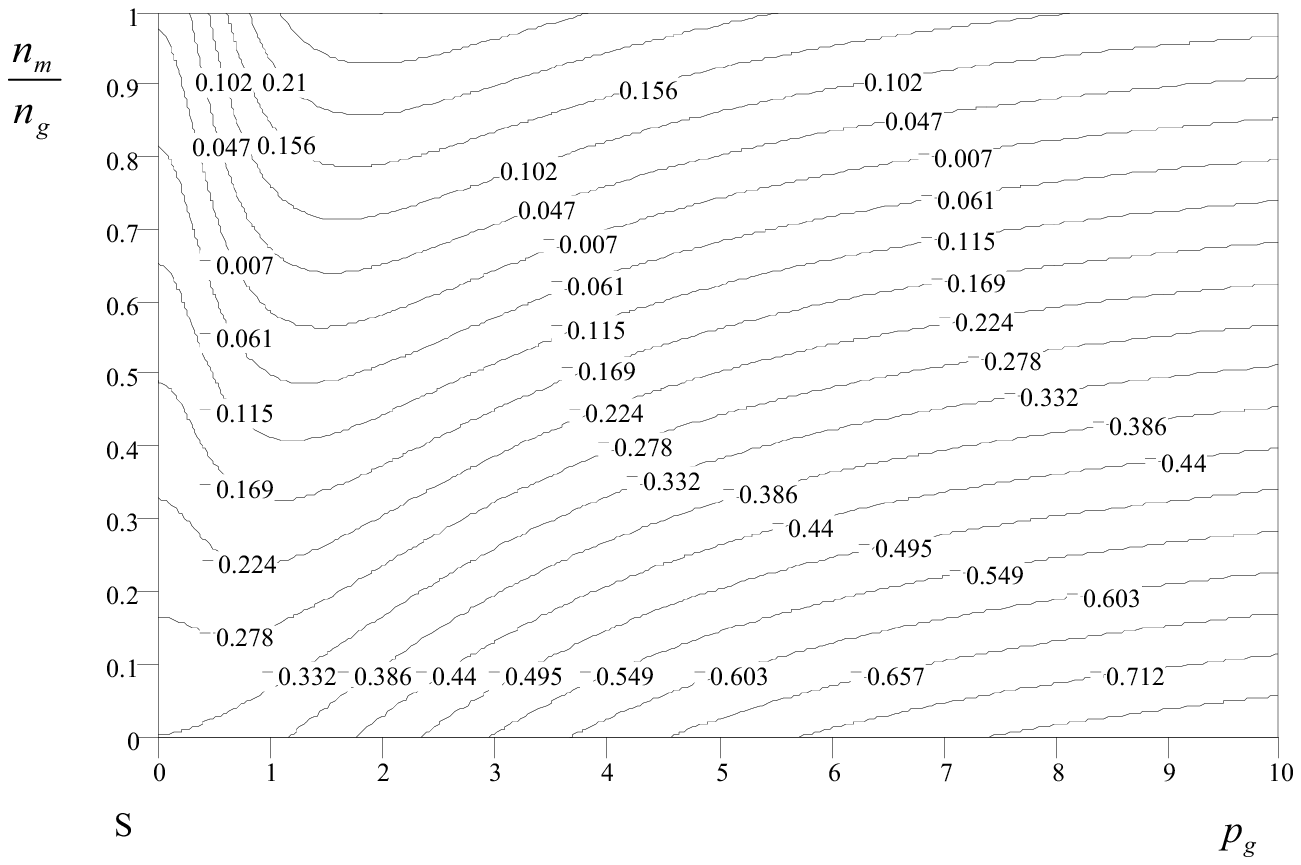}
\caption{\label{f2}Equal values of scaled amplification index
$\alpha'=\alpha/n_g\sigma_0$  at $\mu_m/\mu_g$=4 ($\mathbf{E}
\upuparrows\mathbf{E}_0$).}
\end{center}
\end{figure}
Figure \ref{f1} shows the existence of the optimum value of the
alignment parameter $p_g$ (i.e., $E_0$) for the fixed values
$\mu_m/\mu_g$ and $n_m/n_g$, while Fig. \ref{f2} shows that the
same amplification index can be achieved for different sets of
population and alignment parameters.

In the alternative case, where  the dipole moment in the upper
state is less than in the lower one ($\mu_m<\mu_g$), the alignment
degree in the lower state is larger than for the molecules in the
upper state. Consequently, orthogonal orientation of polarizations
of the probe $\mathbf{E}$ and control $\mathbf{E}_0$ fields
becomes advantageous for the suppression of absorption. Then the
amplification index averaged over the molecule orientation is
given by the equation
\begin{equation}
\alpha=\sigma_0\left\{[n_mL(p_m)/{p_m}]-[n_gL(p_g)/p_g]\right\}.\label{a2}
\end{equation}
\begin{figure}[!h]
\begin{center}
\includegraphics[width=.45\textwidth]{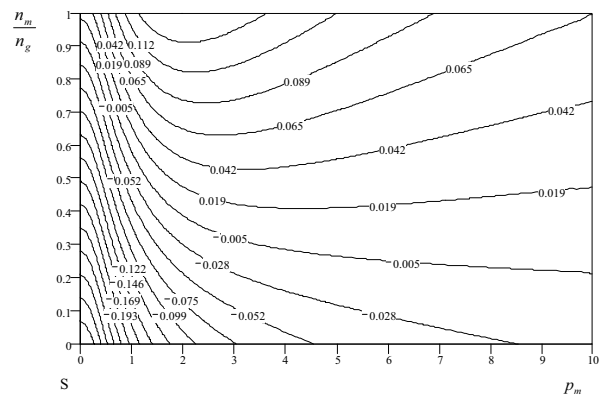}
\caption{\label{f4}Equal values of the scaled amplification index
$\alpha'=\alpha/n_g\sigma_0$  at $\mu_g/\mu_m$=4 ($\mathbf{E}
\perp\mathbf{E}_0$).}
\end{center}
\end{figure}
Note, that contrary to the previous case, AWI becomes possible at
any strength of $E_0$ above the threshold value, which depends on
the specific magnitude of the population ratio $n_m/n_g$
(Fig.~\ref{f3}). Corresponding changes occur in the sets of
parameters attributed to one and the same amplification index
(Fig. \ref{f4}).

Similar conclusions are valid for AWI induced by an ac field
$E_0$, which is stipulated by the difference of the induced dipole
moments in the upper and lower states.

We shall discuss further the suitable types of molecules and
important related features. As an example, the molecule
3-6-Diacetyl-amino-phtalimide can be considered because it does
not possess  permanent dipole moment in the excited state, i.e.,
$\mu_m=0$, while $\mu_g\approx5.5$~D \cite{Ter}. In this case, at
$ p_g\rightarrow\infty$, and $\mathbf{E} \perp\mathbf{E}_0$,
amplification is determined only by the upper-state molecules and
does not depend on population of the lower state:
\begin{equation}
\alpha=\sigma_0n_m/3.\label{ad}
\end{equation}
This is because the lower-state molecules are aligned orthogonal
to the the probe field and, therefore, are not coupled with this
field. On the contrary, the excited molecules are decoupled from
the control field $E_0$, their orientation remains isotropic in
dipole approximation and, hence, the averaged projection of
transition dipole moment $\mathbf{d}_{mg}$ on the probe field
$\mathbf{E}$ is not equal to zero.

The important requirement for the realization of the proposed
technique is that the lifetime in the excited state $\tau$ must
exceed the time $\tau_0$ necessary to set the orientation
equilibrium. This is because the molecules aligned at ground state
must achieve a new distribution, corresponding to the alignment
parameter at the upper state, after excitation to that state. The
characteristic  lifetime for the electrodipole transition is
$\tau\simeq 10^{-8} - 10^{-9}$s. The alignment time may vary in
the wide range depending on the molecule size and on the viscosity
of the solvent. For big organic molecules in the solvents the
characteristic time is $\tau_0\simeq 10^{-10} - 10^{-12}$ s, for
protein macromolecules $\tau_0\simeq 10^{-6} - 10^{-8}$ s, and
$\tau_0\simeq 10^{-2} - 10^{-4}$ s for big biomacromolecules.
Thus, for the applications being discussed here the molecule  mass
must not significantly exceed $10^3$ atomic units and their length
-- some $10 \AA$. For such compounds the constant electric dipole
moment is usually on the order of $1-10$~D \cite{Kie,Ter,Bak}.
Therefore the alignment parameter achieved with the nearly
breakdown dc field $E_0$ can hardly exceed the value about unit,
even in the most electrically resistant solvents \cite{Kik}.
However, the breakdown threshold can be increased by the use of
the pulsed dc fields $E_0$ and by the decrease of  the temperature
of the solvents, which is possible making use of cryogenic liquids
or cooled gas. The alignment parameters above the unit seem to be
achieved by applying a dc field $E_0$ to relatively large
molecules having dipole moment on the order of $10^2 - 10^3$~D.
However, their reorientation time $\tau_0$ most likely is longer
than the lifetime at the exited state. Thus, the effect of AWI,
apparently, may become easier to achieve through orientation of
molecules by ac optical field.

In the case of optical orientation, the polarizability anisotropy
$b_{33} - b_{11}$ is about $10^{-23}$ cm$^3$  even for the
resonance detuning on the order of $|\omega_{ij}-\omega_{0}|\simeq
\omega_{0}$. At normal temperature T=300 K,  optical alignment
parameter $q$  defined by Eq. (\ref{pq}) can reach the magnitude
about unit under focused laser radiation of the power about 10$^6$
W. The increase of intensity of the control field up the breakdown
threshold allows one to achieve the complete optical orientation
of molecules, with the size and weight not exceeding the
magnitudes discussed above. Besides that, the magnitude of $q$ can
be considerably increased by setting the frequency $\omega_0$ in
the vicinity of of the resonant frequencies of the adjacent
optical transitions.

In summary,  we propose to achieve amplification and lasing
without population inversion by anisotropic molecules through
decoupling polarized light from the absorbing molecules at lower
energy level while the upper-state molecules stay emitting. Such
switching becomes feasible with the aid of control electric field
that aligns anisotropic molecules in upper and lower energy levels
in a different way. The example of suitable molecules similar to
dye molecules is given and the favorable conditions are discussed
and illustrated. The achievable magnitudes of amplification
without population inversion are estimated based on the derived
equations.

We thank V. V. Laschinsky for help with computing and drawing the
graphs.

\end{document}